\documentclass{ws-procs9x6}

\newcommand{\ltsima}{$\; \buildrel < \over \sim \;$}
\newcommand{\simlt}{\lower.5ex\hbox{\ltsima}}
\newcommand{\gtsima}{$\; \buildrel > \over \sim \;$}
\newcommand{\simgt}{\lower.5ex\hbox{\gtsima}}

\newcommand{\pn}{\par\noindent}

\def\lesssim{\mathrel{\hbox{\rlap{\hbox{\lower4pt\hbox{$\sim$}}}\hbox{$<$}}}}
\def\gtrsim{\mathrel{\hbox{\rlap{\hbox{\lower4pt\hbox{$\sim$}}}\hbox{$>$}}}}

\def\ab1450{$AB_{1450(1+z)}$}

\def\xray{\hbox{X-ray}}

\def\today{\ifcase\month\or January\or February\or March\or April\or May\or
      June\or July\or August\or September\or October\or November\or December\fi
      \space\number\day, \number\year}
\def\chandra{{\it Chandra\/}}

\def\heao1{{\it HEAO-1\/}}

\def\rosat{{\it ROSAT\/}}

\def\xmm{{XMM-{\it Newton\/}}}

\def\apj{{\it ApJ}}

\def\aap{{\it A\&A}}
\def\aj{{\it AJ}}

\def\etal{{\it et al.}}
\begin{document}

%\title{X-ray emission from high-redshift quasars: a \chandra\ and \xmm\ 
%update}
\title{
%News from outer space: \\ 
The \boldmath$z>4$ quasar population observed by Chandra and XMM-Newton}

\author{C. VIGNALI\footnote{\uppercase{W}ork partially supported by the 
\uppercase{I}talian \uppercase{S}pace \uppercase{A}gency under the contract 
\uppercase{ASI} \uppercase{I/R}/057/02 \uppercase{(CV)} 
and by the \uppercase{C}handra \uppercase{X}-ray 
\uppercase{C}enter grant \uppercase{GO}2-3134\uppercase{X} \uppercase{(WNB)}.}}

\address{INAF--Osservatorio Astronomico di Bologna, \\
Via Ranzani, 1 -- I-40127 Bologna, Italy \\
%E-mail: l\_vignali@bo.astro.it}
E-mail: cristian.vignali@bo.astro.it}

\author{W.N. BRANDT and D.P. SCHNEIDER}

\address{Department of Astronomy \& Astrophysics, \\ 
The Pennsylvania State University, \\ 
525 Davey Lab, University Park, PA 16802, USA \\
E-mail: niel,dps@astro.psu.edu}


\maketitle

\abstracts{
The current status of our \chandra\ and \xmm\ project on 
high-redshift ($z\simgt4$) quasars is briefly reviewed. 
We report the main results obtained in the last 
few years for the \xray\ detected quasars, along with 
a few ($\approx$~10\%) intriguing cases where no 
detection has been obtained with \chandra\ snapshot observations.
}

\section{Introduction}
Quasars at $z\simgt4$ provide direct information on the first massive 
structures to form in the Universe. The last few years have evidenced 
incredible progress in the study of the high-redshift 
Universe, largely thanks to ground-based optical surveys. 
In particular, the Sloan Digital Sky Survey (SDSS\cite{y00}) 
has discovered large numbers of high-redshift quasars, increasing the 
number of known quasars at $z\simgt4$ to $>500$,\cite{f04} 
and many more are expected to be discovered 
before the survey ends (see [3] 
%will come to an end\cite{ab04} 
for the recent advances provided by the SDSS). 
Many of the high-redshift quasars discovered by recent optical surveys 
are suitable for follow-up \xray\ investigations. Here we review \xray\ 
studies of the highest redshift quasars, focusing on recent advances enabled 
largely by the imaging and spectroscopic capabilities of \chandra\ and \xmm.

%\section{X-ray results}
\section{X-ray detections at $z\simgt4$}
Our knowledge of the \xray\ properties of $z\simgt4$ quasars has advanced 
rapidly over the last few years. Since the pioneering work based on the 
systematic analysis of archival \rosat\ data\cite{kbs00}, where the number 
of \xray\ detected quasars at $z>4$ doubled, the progress made in this field 
has been substantial (see Fig.~1). 
At present, more than 90 quasars at $z\simgt4$ have \xray\ 
detections;\footnote{See 
http://www.astro.psu.edu/users/niel/papers/highz-xray-detected.dat for a 
regularly updated listing of \xray\ detections and sensitive upper limits 
at $z\simgt4$.} while in most cases \chandra\ snapshot 
(typically \hbox{4--6~ks}) observations have allowed derivation of 
basic \xray\ information (\xray\ fluxes and 
luminosities),\cite{b01,v01,b02,sil02,be03,v03a,v03b,cas03,tre03} 
for a few \xray\ luminous objects medium-quality \xray\ spectra 
have also been obtained\cite{fb03,gr04} thanks to the large collecting area 
of \xmm. 
%
%%%%%%%%%%%%%%%%%%%%%%%%%%%%%%%%%%%%%%%%%%%%%%%%%%%%%%%%%%%%%%%%%%%%%%
% Figure 1 - two panels: (a) Fx vs. AB1450(1+z) just after KBS00 work
%                        (b) Fx vs. AB1450(1+z): present situation
\begin{figure}[t]
\parbox{0.48\textwidth}
{\psfig{figure=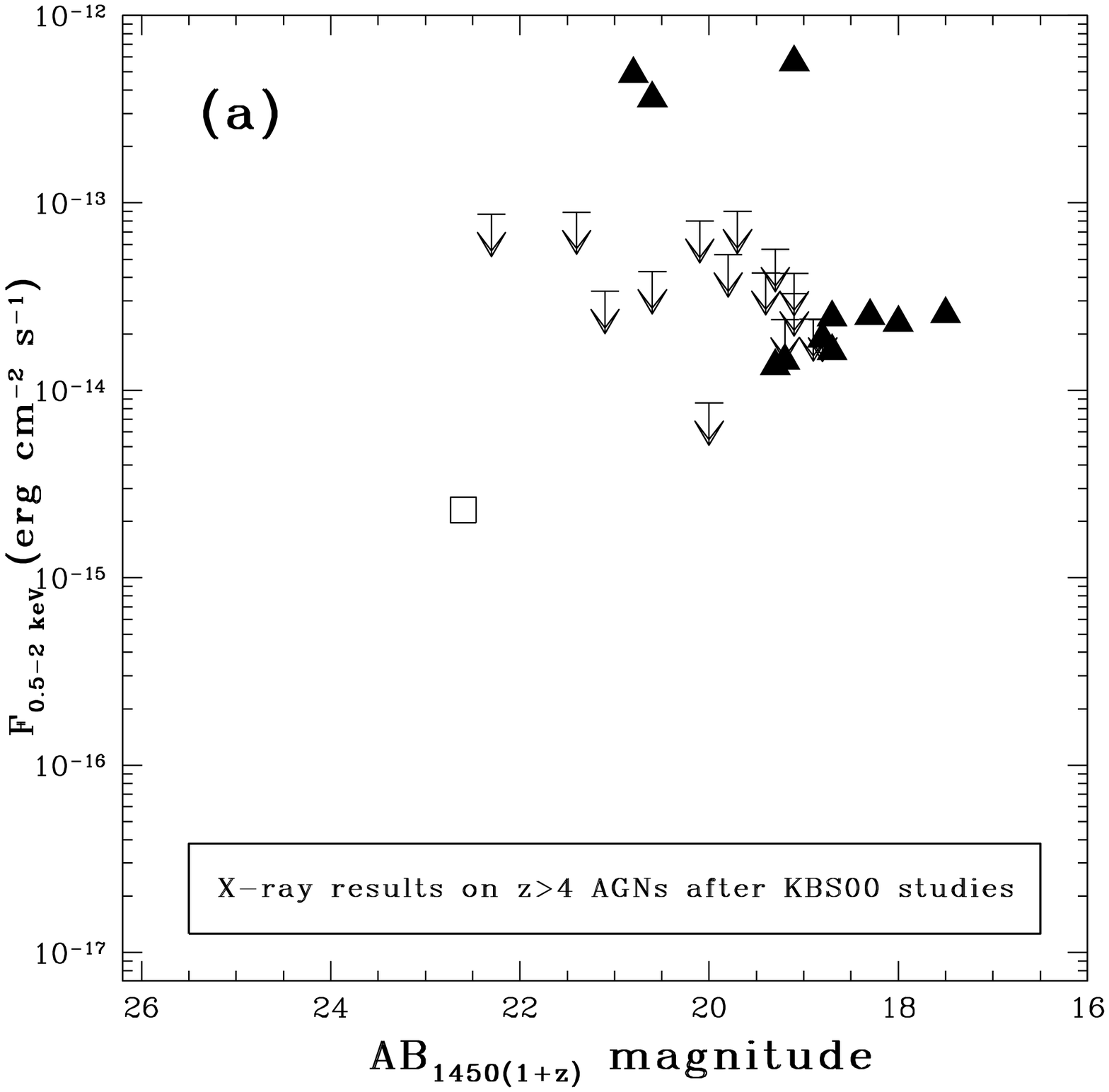,width=14pc,angle=0}}
\hglue-0.8cm 
% \hglue=-0.8cm if width=14pc and fxab_now_nolabel.eps is used
% otherwise: USE no space, width=13pc, and fxab_now,eps 
\parbox{0.48\textwidth}
{\psfig{figure=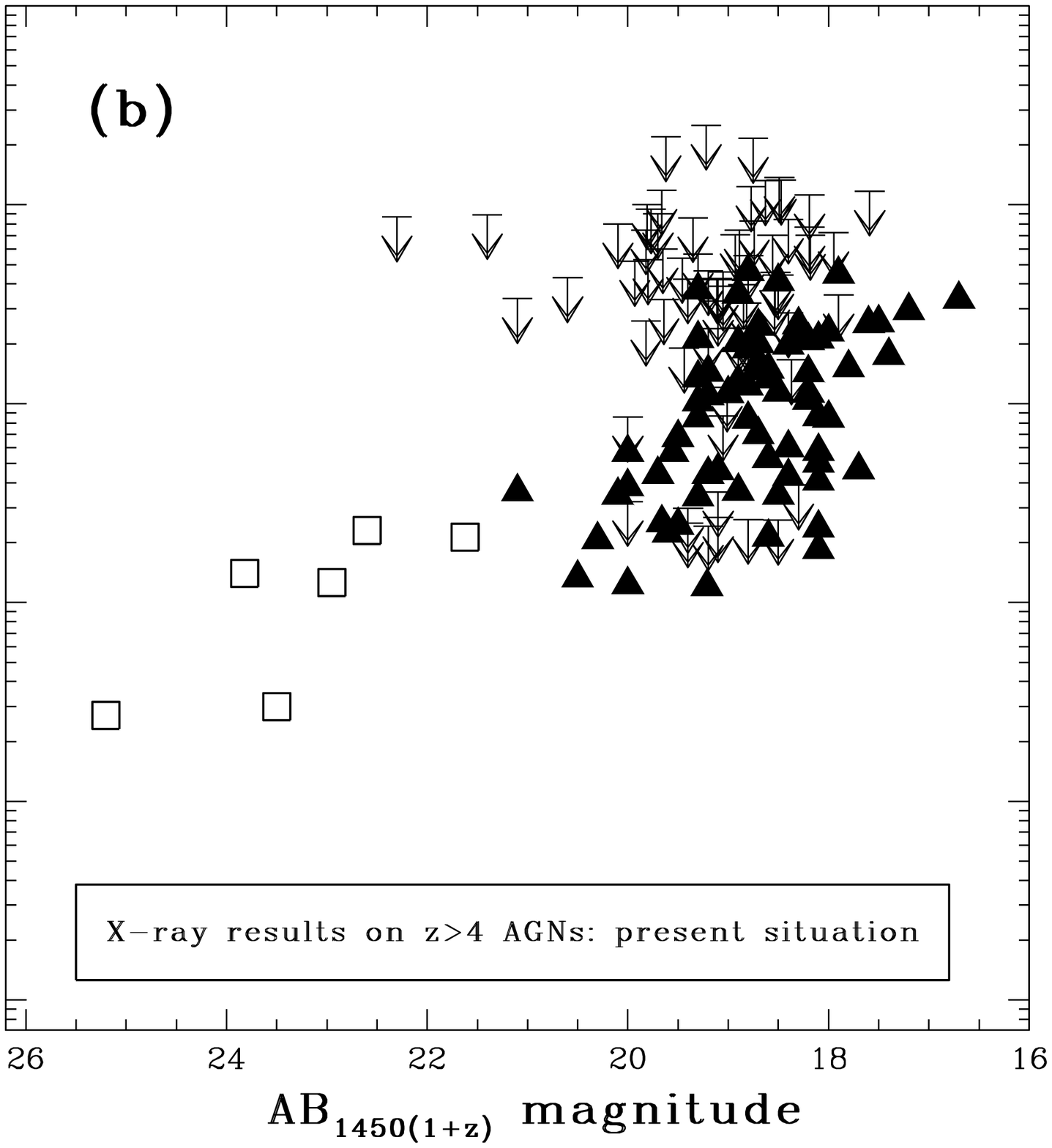,width=14pc,angle=0}}
\caption{Observed-frame, Galactic absorption-corrected 0.5--2~keV flux versus 
\ab1450\ magnitude for $z\simgt4$ AGN found in optical (triangles) and 
\xray\ (squares) surveys. The rapid increase in the number of \xray\ 
detected AGN at $z\simgt4$ is evident when the two panels (i.e., the 
situation in year 2000 vs. the current one) are compared. 
Most of the upper limits have 
been obtained from archival \rosat\ observations.}%\cite{kbs00,v01,v03a}
\label{fig1}
\end{figure}
%%%%%%%%%%%%%%%%%%%%%%%%%%%%%%%%%%%%%%%%%%%%%%%%%%%%%%%%%%%%%%%%%%%%%%
%
From a general perspective, the most important results from \xray\ 
analyses of high-redshift quasars are the following: 
%%%%%%%%%%%%%%%%%%%%%%%%%%%%%%%%%%%%%%%%%%%%%%
% Summary of main X-ray results on high-z QSOs
%%%%%%%%%%%%%%%%%%%%%%%%%%%%%%%%%%%%%%%%%%%%%%
\pn
$\diamond$ The \xray\ properties of $z\simgt4$ quasars are similar to 
those of local quasars. Joint \xray\ spectral fitting of optically 
luminous quasars at $z\simgt4$ selected from the Palomar Digital Sky Survey 
(DPOSS\cite{dj98}) and the SDSS has shown\cite{v03a,v03b} that 
quasar \xray\ continuum shapes 
do not show evidence for significant spectral evolution over cosmic time.
This has found further support recently via joint spectral fitting of a 
sample of 46 radio-quiet quasars (RQQs) with \chandra\ detections 
in the redshift range 4.0--6.3 ($\approx$~750 source counts; 
\hbox{$\Gamma=1.9\pm{0.1}$})\cite{vbs04} and by direct 
\xray\ spectral analyses of a few objects observed by \xmm.\cite{fb03,gr04}
No evidence for widespread \xray\ absorption has been found, 
although some quasars (see discussion below) are likely to be obscured. 
\pn 
$\diamond$ The optical-to-\xray\ properties of high-redshift 
quasars and those of local quasars have been found similar, once 
luminosity effects and selection biases are taken 
into account properly.\cite{vbs03}
%%
%Significant differences between the high-redshift 
%quasar population and the local one have been found in their broad-band 
%properties using the SDSS Early Data Release quasar catalog;\cite{dps02} 
%detailed investigations on a well-selected and homogeneous sample 
%have confirmed previous results indicating that 
%these difference are due to luminosity effects.\cite{vbs03}
%%
%\pn
%$\diamond$ \xray\ spectral studies do not show hints for different accretion 
%and growth mechanisms, such as accretion-disk instabilities 
%and trapping-radius effects. 
%\pn
Overall, the emerging picture is that the small-scale \xray\ emission 
regions of quasars appear relatively insensitive to large-scale environmental 
differences at $z\approx$~6. 
\pn 
$\diamond$ Moderately deep \chandra\ observations and the ultra-deep (2~Ms) 
survey in the \chandra\ Deep Field-North (CDF-N)\cite{ale03} have 
allowed \xray\ studies of moderate-luminosity AGN at $z>4$; 
this population of AGN is more numerous and thus more 
representative than the rare, highly luminous quasars.\cite{v02}
\pn
$\diamond$ Similarly to the results obtained for the RQQs, neither the 
\xray\ spectral slope ($\Gamma=1.65\pm{0.15}$) nor the jet emission 
of the radio-loud quasar population seems to evolve with 
cosmic time.\cite{bas04} 
No evidence for significant \xray\ brightening ascribed to 
inverse Compton scattering of energetic electrons 
with Cosmic Microwave Background photons 
has been revealed by snapshot observations with \chandra.\cite{bas04}

\section{X-ray non-detections: space oddities at high redshift?}
At present, the fraction of \xray\ upper limits among the sensitive \chandra\ 
and \xmm\ observations of high-redshift quasars is low ($\approx$~10\%). 
About half of the \xray\ undetected quasars are broad 
absorption-line quasars (BALQSOs), which are known to be absorbed 
in the \xray\ band;\cite{blw00} one quasar shows narrow absorption 
features from N\,{\sc v} and Ly$\alpha$ at approximately the 
source redshift.\cite{v03b} 
For most of these objects the lack of \xray\ detections at the flux 
limits reached by \chandra\ with snapshot observations is likely due to 
absorption by column densities larger\cite{v01,v03b} 
than $\approx$~10$^{23}$~cm$^{-2}$ 
(assuming reasonable values for the optical-to-\xray\ spectral energy 
distribution and \xray\ photon index). 
It is possible that the remaining \xray\ undetected high-redshift quasars 
are characterized by absorption features not recognized in low-resolution 
optical spectra. Recently, it has been shown that 
near-infrared spectroscopy can be highly effective\cite{ma04} 
in detecting absorption features in high-redshift objects 
previously classified as ``normal'' quasars. 
Clearly, minority AGN populations at $z>4$ need to be investigated better 
in the \xray\ regime. These include 
%moderate-luminosity AGN, 
BALQSOs (whose number is still too limited to derive statistically 
reliable average properties) and the population of quasars 
lacking emission lines.\cite{fan99,v01} 
Follow-up observations of quasars without optical emission lines 
with \chandra\ are on-going and the results will be presented in a 
forthcoming paper.\cite{br04}


\begin{thebibliography}{0}
%%%
\bibitem{y00} 
D.G. York \etal, \aj\ {\bf 120}, 1579 (2000). 
%
\bibitem{f04}
X. Fan, private communication (2004). 
%
\bibitem{ab04} 
K. Abazajian \etal, \aj, in press, astro-ph/0403325 (2004). 
%
\bibitem{kbs00} 
S. Kaspi, W.N. Brandt and D.P. Schneider, \aj\ {\bf 119}, 2031 (2000; KBS00). 
%
\bibitem{b01} 
%W.N. Brandt, M. Guainazzi, S. Kaspi, X. Fan, D.P. Schneider, 
%M.A. Strauss, J. Clavel and J.E. Gunn, \aj\ {\bf 121}, 591 (2001). 
W.N. Brandt \etal, \aj\ {\bf 121}, 591 (2001). 
%
\bibitem{v01}
C. Vignali, W.N. Brandt, X. Fan, J.E. Gunn, S. Kaspi, D.P. Schneider and 
M.A. Strauss, \aj\ {\bf 122}, 2143 (2001).
%
\bibitem{b02} 
W.N. Brandt \etal, \apj\ {\bf 569}, L5 (2002).
%
\bibitem{sil02}
J.D. Silverman \etal, \apj\ {\bf 569}, L1 (2002). 
%
\bibitem{be03} 
J. Bechtold \etal, \apj\ {\bf 588}, 119 (2003).
%
\bibitem{v03a}
C. Vignali, W.N. Brandt, D.P. Schneider, G.P. Garmire and S. Kaspi, 
\aj\ {\bf 125}, 418 (2003). 
%
\bibitem{v03b}
C. Vignali \etal, \aj\ {\bf 125}, 2876 (2003).
%
\bibitem{cas03} 
F.J. Castander, E. Treister, T.J. Maccarone, P.S. Coppi, 
J. Maza, S.E. Zepf and R. Guzman, \aj\ {\bf 125}, 1689 (2003).
%
\bibitem{tre03} 
E. Treister, F.J. Castander, T.J. Maccarone, D. Herrera, E. Gawiser, J. Maza 
and P.S. Coppi, \apj\ {\bf 603}, 36 (2004).
%
\bibitem{fb03}
E. Ferrero and W. Brinkmann, \aap\ {\bf 402}, 465 (2003). 
%
\bibitem{gr04}
D. Grupe, S. Mathur, B.J. Wilkes and M. Elvis, \aj\ {\bf 127}, 1 (2004).
%
\bibitem{dj98}
S.G. Djorgovski \etal, in Proceedings of ``Wide Field Surveys in Cosmology'', 
%ed. S. Colombi, \& Y. Mellier (Paris: Editions Frontieres), 
89 (1998). 
%
\bibitem{vbs04}
C. Vignali, W.N. Brandt and D.P. Schneider, in Proceedings of 
``AGN Physics with the Sloan Digital Sky Survey'', in press, 
astro-ph/0310659 (2004). 
%
\bibitem{vbs03}
C. Vignali, W.N. Brandt and D.P. Schneider, \aj\ {\bf 125}, 433 (2003). 
%
\bibitem{ale03}
D.M. Alexander \etal, \aj\ {\bf 126}, 539 (2003). 
%
\bibitem{v02}
C. Vignali, F.E. Bauer, D.M. Alexander, W.N. Brandt, A.E. Hornschemeier, 
D.P. Schneider and G.P. Garmire, \apj\ {\bf 580}, L105 (2002).
%
%\bibitem{dps02}D.P. Schneider \etal, \aj\ {\bf 123}, 567 (2002). 
%
\bibitem{bas04}
L.C. Bassett, W.N. Brandt, D.P. Schneider, C. Vignali and G.P. Garmire, 
\aj, in press, astro-ph/0404543 (2004). 
%
\bibitem{blw00}
W.N. Brandt, A. Laor and B.J. Wills, \apj\ {\bf 528}, 637 (2000). 
%
\bibitem{ma04}
R. Maiolino, E. Oliva, F. Ghinassi, M. Pedani, F. Mannucci, R. Mujica and 
Y. Juarez, \aap, in press, astro-ph/0312402 (2004). 
%
%\bibitem{and01} S.F. Anderson \etal, \aj\ {\bf 122}, 503.
\bibitem{fan99}
X. Fan \etal, \apj\ {\bf 526}, L57 (1999).
%
\bibitem{br04}
W.N. Brandt \etal, in preparation. 
%%%
\end{thebibliography}
\end{document}